# Photochargeable Li-ion Battery: Type II Heterojunction Exposes Underlying Band Gap in 'Metallic' Lithiated MoS$_2$

*Raheel Hammad, Amar Kumar, Tharangattu N. Narayanan\*, and Soumya Ghosh\**

**Abstract:**
Light chargeable metal-ion batteries using semiconductor heterostructures are gaining enormous interest. A few such heterostructures such as MoS$_2$/MoO$_y$ and TiS$_2$/TiO$_2$ have been shown to function as photocathodes in photochargeable Li-ion batteries, where the type II set-up has been proposed to generate spatially separated (longer-lived) excitons upon photo-exposure. The Li intercalated MoS$_2$, generated during the discharge cycle of the battery, undergoes a phase transition from the semiconducting (2H) to a metallic (1T') phase, in contrast to its TiS$_2$ counterpart, casting a doubt over the photocharging process. Here, employing density functional theory based traditional as well as unconventional computational schemes along with relevant spectroscopic techniques, we show that in Li$_x$MoS$_2$/MoO$_3$ heterostructure an underlying band gap of Li$_x$MoS$_2$ is exposed, upto a certain value of x, due to dispersion of electron density onto MoO$_3$ justifying the observed photocharging. We believe that the general concepts explored in this study will be important in the rational design of photo-cathode materials in Li-ion batteries.

**TOC graphic:**

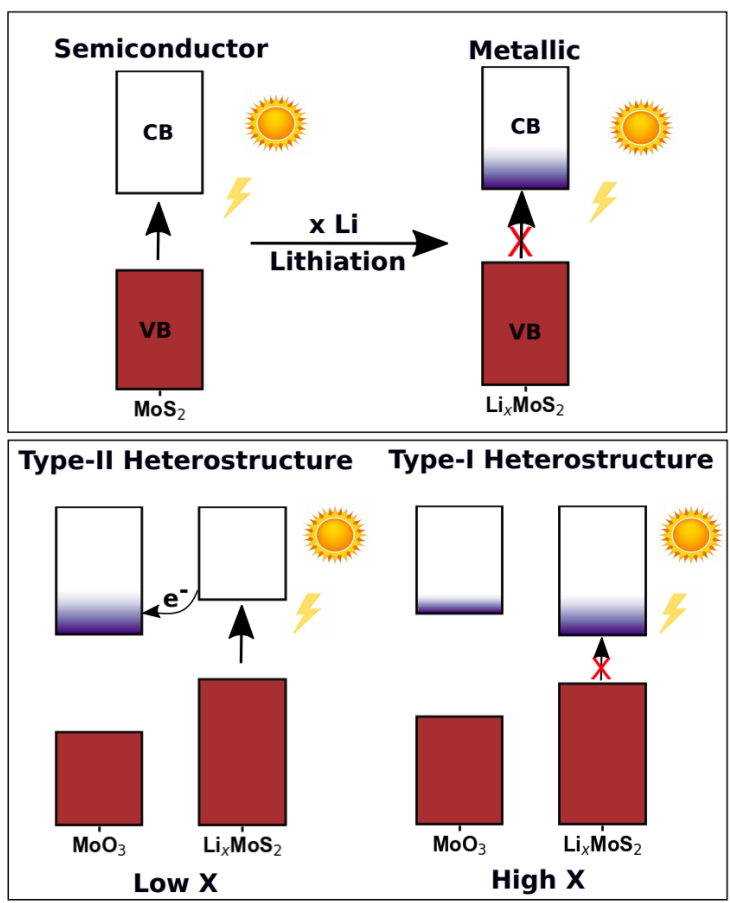

**Introduction**

Following the successful demonstration of LiCoO$_2$ as a viable cathode material by Goodenough and Mizushima,[1] lithium ion batteries have revolutionized the portable electronic industry. In recent times, layered transition metal oxide based LiNi$_x$Mn$_y$Co$_z$O$_2$ (x + y + z = 1), commonly termed as NMC, class of structures have shown immense potential as cathode materials.[2–5] These Ni-based layered oxides where x < 0.8 are currently in use in EVs.[6–9] Increasing the Ni-content beyond 0.33 closes the band gap gradually indicating that the Ni-rich complexes are metallic. Concomitantly, there is also an enhancement in the overall electrical conductivity due to increase in the charge carrier density owing to the presence of higher concentration of Ni$^{3+}$ ions.[10,11] But there is still room for further improvement. Increasing the Ni-content leads to higher capacity at the expense of cyclability. The decrease in structural stability at high Ni-content has been attributed to abrupt phase transformations in nano-domains and lower surface energy.[12] Recently, several studies have been directed towards understanding the degradation mechanism of NMC based electrochemical cells.[13,14] Several attempts have been made to stabilize the structure with dopants or coat the surface of the active material with other less reactive metal oxides to suppress any parasitic reactions.[15–17]

An alternative strategy is to look for cathode materials that can be charged with the help of an external light source, which might bypass some of the issues plaguing the traditional batteries. In addition to maintaining high capacity, the photo-cathode materials should have a band gap commensurate with the frequency of light in the visible region. Early examples on photochargeable materials involved an integration of multiple components, one for photoexcitation and another for Li-ion adsorption.[18,19] In 2018, Ahmed *et al* reported a single 2D perovskite material for both light absorption and ion storage.[20] On the other hand, a type II heterojunction has been employed for photosensing and photocatalytic applications in order to efficiently separate the charge carriers that are generated upon photo-excitation.[21–23] This concept has been recently extended to photocathodes.[24–27] The efficiency of the type II set-up is critically dependent on the proper alignment of the band edges of the two materials that ensures dispersion of the excited electron on the conduction band (CB) of one material to the CB of the other.[24–28] Hence, for a type II heterojunction to act as a photocathode, the bands of the two materials need to remain properly aligned *during the full discharge cycle of the battery.* While MoS$_2$/MoO$_y$ has been recently demonstrated in a photorechargeable Li ion battery,[27] previous studies indicate that upon lithiation, MoS$_2$ becomes metallic irrespective of the initial phase (Figure S1).[29,30] The metallization questions the photocharging mechanism, where exciton formation has to happen upon the absorption of light in the lithiated MoS$_2$ (Li$_x$MoS$_2$) (Scheme 1).[27] In this study, we demonstrate how lithiated MoS$_2$ can still be photoexcited upto a certain concentration of intercalated lithium ion in conjunction with MoO$_3$.

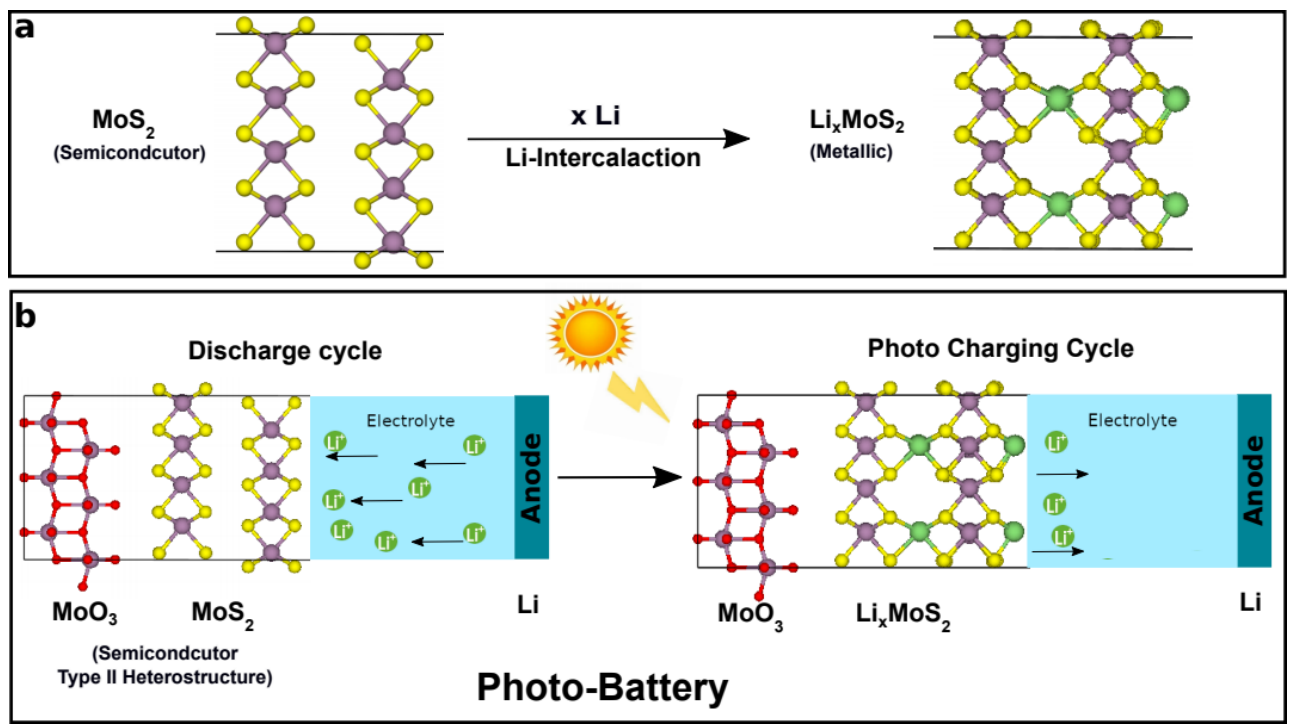

**Scheme 1.** Schematic representation illustrating the photocharging of Li$_x$MoS$_2$, an apparently metallic material, in the Li-ion battery.

## Methods
### Computational Details

Our calculations are based on Density functional theory (DFT) which uses PBE form for the Generalized Gradient approximation(GGA) and the hybrid functional HSE06.[31–33] These calculations were performed in the Vienna ab-initio simulation package(VASP) using a plane wave basis set with the projector augmented plane-wave method (PAW).[34,35] For all MoS2 calculations a kinetic energy cutoff for plane-waves was set to be 360 eV, while a cutoff of 500 eV was used for $MoO_3$ and the $MoO_3/MoS_2$ heterojunction calculations. The energy and force convergence criteria were set to be 10 $^{-4}$ eV and 0.02 eV Å$^{-1}$ respectively. The Gamma Centered Monkhorst scheme of grid density 0.025 (2 π Å −1 ) in each direction was used to sample the Brillouin zone for all the calculations. As pointed out by Yan Hua Lei et.al accurate MoO3 calculations require Hubbard correction on top of DFT.[36] Therefore we have used DFT+U calculations introduced by Dudarev et. al. with a hubbard parameter of 5.0 eV for both MoS2 and $MoO_3$.[36,37] For all multi Layered systems DFT+U was combined with the Grimes et.al dispersion correction(D3) to account for the long range interaction.[38]

We used planar averaged Hartree potential to compute the offset between $MoO_3$ and $MoS_2$. The valence band maxima(VBM) of bi-layer $MoS_2$(2L) is computed with HSE06 functional relative to its macroscopic average, which in turn is referenced to the vacuum level in a slab calculation. For bulk $MoO_3$ calculations (with HSE06), the VBM is referenced to the vacuum level using the following protocol. We first computed the electronic structure of multi-layered slabs of $MoO_3$ [010] in conjunction with vacuum in addition to the bulk $MoO_3$. The lateral dimensions of the slab and the bulk simulation cells are kept the same while the geometry was relaxed in both cases. Central layers in the slab are supposed to represent the bulk and using the macroscopic average of this region one can reference the VBM of the bulk to the vacuum.[39,40] We can then compute the valence band offset (VBO) between the two materials. The simplified formula for the above method is given below

$$\text{VBO} = \widetilde{VBM}_{2MoS_2} + \overline{\overline{V}}_{MoO_2-vac} - \left(VBM_{MoO_3} - \overline{\overline{V}}_{MoO_3} + \overline{\overline{V}}_{MoO_3} - \overline{\overline{V}}_{vac}\right) \quad (1)$$

$$= \widetilde{VBM}_{2MoS_2} + \overline{\overline{V}}_{2MoS_2-vac} - \left(\widetilde{VBM}_{MoO_3} + \overline{V}_{MoO_3-vac}\right)$$

$\overline{\overline{V}}_x =$ macroscopic average;

$\overline{\overline{V}}_{x-vac} =$ macroscopic average of the bulk referenced to the vacuum;

$VBM_x =$ VBM with respect to the macroscopic average

The conduction band offset is simply obtained by adding the band-gap to the VBM. However, PBE is known to underestimate the band gap, and therefore the experimental/HSE06 band gap can be combined with the above VBO for band alignment.

### Experimental section

Chemical vapor deposition (CVD) technique is employed for the development of $MoS_2$ monolayers. A detailed synthesis procedure is discussed in our previous work, [41] and also explained in the supporting information. Lithium ion battery half-cell is constructed using these monolayers transferred on an ITO coated quartz plate (25 mm X 25 mm) and lithium foil as another electrode with LiPF6- ethylene carbonate and ethyl methyl carbonate electrolyte. A customized home-built battery assembly system having a light entering window is used for battery assembly.[26,27] In situ Raman and photoluminescence (PL) studies are conducted using a micro-Raman spectrometer with 532 nm excitation, details in supporting information.

### Results and Discussion

We begin our studies by exploring the band structure of $MoS_2$ and then investigate the alignment of bands between bilayer $MoS_2$ and bulk $MoO_3$. Bulk $MoS_2$ in 2H phase (with 2 layers of hexagonal lattice stacked in AB fashion) displays an indirect band gap of ~ 1.3 eV while the 1T' phase (monolayer of distorted octahedral phase) is found to be metallic.[42–44] The band gap of H-$MoS_2$ increase on going from bulk to monolayer[42] while there is a small indirect band gap in the computed band structure of 1T' monolayer.[42] In order to understand the electronic structure of the $MoS_2$ (bilayer)[45]/$MoO_3$ (bulk) heterostructure, one needs to align the band edges between the two materials that requires computation of the valence band maxima (VBM) of the two materials relative to the vacuum in addition to their band gaps, which invariably involves calculations with finite number of slabs as explained in the computational details section.[46] If both the bulk and the slab structures for $MoO_3$ are built with the conventional layers (Figure 1a), then the composite band structure with bilayer H-MoS2 displays a type-III behavior (Figure 1b) when PBE+U functional is employed for the calculations, mainly because the band gaps are severely underestimated. In order to obtain a type-II alignment, Figure 1c, with the same structures one would need to compute the band gaps with a more accurate hybrid functional (HSE06) while using PBE+U+D3[37,38] for all the other relevant quantities to estimate the valence band offset (VBO).

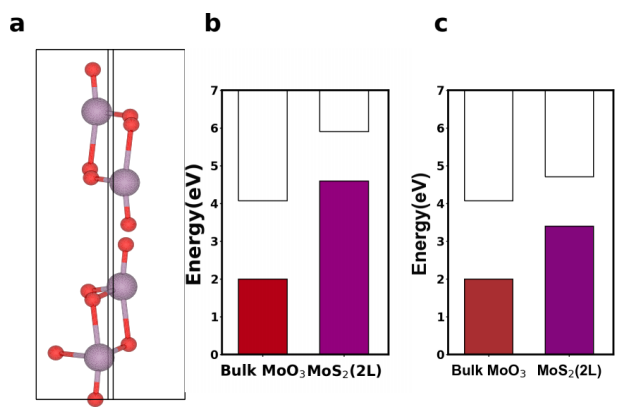

**Figure 1**. (a) Unit cell for conventional layering (b) the band alignment between bilayer 2H-MoS$_2$ and bulk MoO$_3$ where both the VBO and the band gap is computed with PBE+U  c)  the band gap is computed with HSE06 but the VBO is computed with PBE+U

As pointed out earlier (Scheme 1), during the discharge cycle of the battery the exposed cathode material (MoS$_2$) gets lithiated and lithiated MoS$_2$ (Li$_x$MoS$_2$) is metallic. Now, in order to investigate the electron distribution in lithiated MoS$_2$ in conjunction with MoO$_3$, we set-up an explicit heterostructure composed of several units of MoS$_2$ and MoO$_3$, as shown in Figure 2. Note, however, that a comparatively large supercell is required to minimize lattice strain in the heterostructure, and hence, one would need to use GGA functional for practical computational purposes. Incidentally, previous DFT calculations indicated that layered MoS$_2$/MoO$_3$ combination forms a type III heterojunction, analogous to the situation shown in Figure 1b, if PBE functional is employed.[47] Chemical modification of MoO$_3$ layers were previously employed to attain a type-II character.[47]

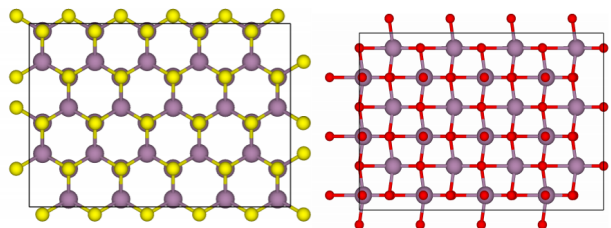

**Figure 2**. Units of MoS$_2$ [5x2](a) and MoO$_3$[4x3](b) employed in the explicit heterostructure calculations

In this work, we introduce an alternative approach. We demonstrate that the type-II alignment can also be obtained by employing a different, albeit artificial, layering scheme for MoO$_3$, termed as 'uL scheme' (Figure 3a), where one considers the layers that are farther apart as one unit. The change in the energy of the band edges on going from a bilayer to bulk is minimal for this layering scheme (Figure S2). The band alignment between bulk MoO$_3$ and bilayer MoS$_2$ is demonstrated in Figure 3b, where the reference slab calculations for MoO$_3$ employed uL scheme and PBE+U+D3 functional is employed for all the components (equation 1 above). A comparison between Figures 1b (natural layering) and 3b (uL scheme for layering) suggests that the band gap of MoO$_3$ remains more or less the same but the alignment of the band edges (VBM and CBM) with respect to bilayer MoS$_2$ has changed, leading to a type-II arrangement (Figure 3b)) instead of type-III (Figure 1b).

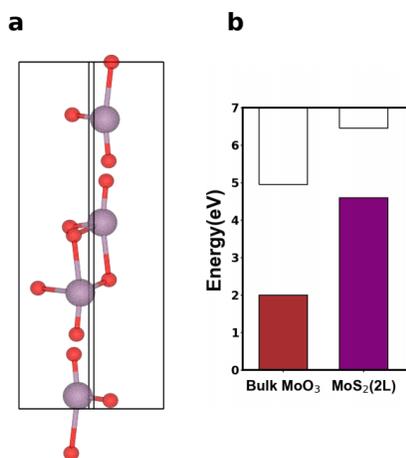

**Figure 3**. Unit cell for uL scheme. Band alignment between bilayer 2H-MoS$_2$ and bulk MoO$_3$ where both VBO and band gap are computed with PBE+U.

In order to investigate the electron distribution between lithiated MoS$_2$ and MoO$_3$, we constructed a supercell for MoS$_2$/MoO$_3$ heterostructure using this uL scheme for MoO$_3$ (Figure 4a). From the corresponding PDOS (Figure 4b) it is evident that even with only 2 layers of both MoS$_2$ and MoO$_3$, we can still obtain a type-II alignment in the *explicit heterostructure* if the uL scheme is employed for MoO$_3$. When two Li-atoms are added to the upper layer (Figure 4c), then a significant amount of charge density (56 %) is delocalized onto the CB of MoO$_3$. The corresponding PDOS is provided in Figure 4d.

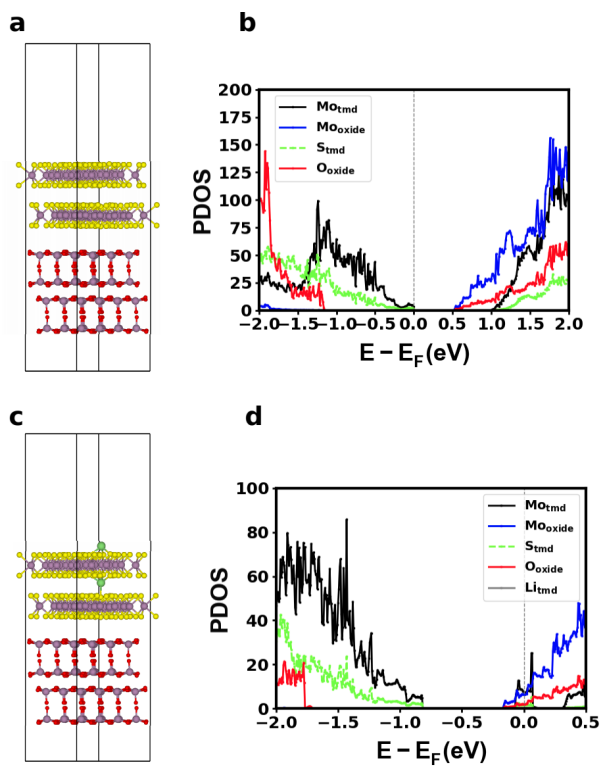

**Figure 4.** (a) Supercell for 2H-MoS$_2$/MoO$_3$ (uL) heterostructure (b) Corresponding PDOS resolved into Mo (MoS$_2$, black), Mo (MoO$_3$, blue), S (green), O (red) density of states. The heterostructure exhibits band gaps of 1.0 eV and 2.2 eV for MoS$_2$ and MoO$_3$, respectively along with a type II character since the CB minimum of MoO$_3$ lies inside the band gap of MoS$_2$. (c) Supercell of the explicit heterostructure with 2 Li-ions (d) Corresponding PDOS projected separately on Mo(MoS$_2$), Mo(MoO$_3$), O, S, Li.

The above sections demonstrate that the heterostructure is able to extract electron density from the CB of lithiated MoS$_2$ that will expose an underlying band gap. As pointed out in the introduction, an ideal photocathode should maintain a robust band gap throughout the complete discharge cycle. In order to check whether this is true for lithiated MoS$_2$ in conjunction with bulk MoO$_3$, we computed the band alignment as a function of increasing concentration of intercalated Li-ion for the two schemes described above (Figure 1c and Figure 3b). The results are shown In Figure 5,[47,48] along with an estimate of the number of electrons that is present in the CB of Li$_x$MoS$_2$. As can be clearly seen, the number of electrons in the CB is equal to the number of Li-ions that are added to the system. In principle, if the CB of Li$_x$MoS$_2$ remains at a higher energy as compared to the CB of MoO$_3$ then these electrons can be extracted by MoO$_3$. However, it is clear from the plots in Figure 5 that with increase in the lithium ion concentration there is a dramatic stabilization of the CBM of Li$_x$MoS$_2$ while the changes in the VBM is comparatively smaller. Ultimately, the type II character is lost and hence, there cannot be any efficient photo-excitation beyond that point (x = 0.33, Figure 5a; x = 0.25, Figure 5b), even if an underlying gap remains. Further analysis of the PDOS (Figure 5c-f) shows that the metallization happens due to shifts in the band edges of the top MoS$_2$ layer where Li-atoms have been added.

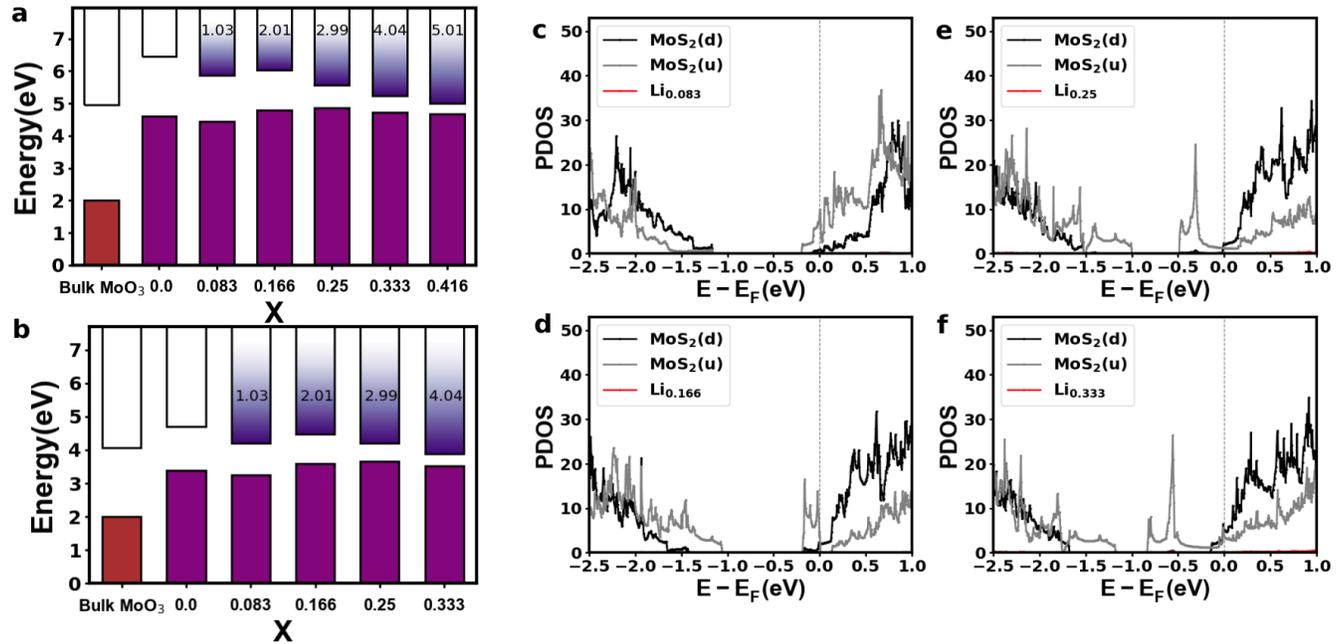

**Figure 5**. Band edges of Li$_x$MoS$_2$ for different values of x with the conventional and uL layering schemes as shown in Figures 1c and 3b, respectively. The type-II character is lost for x > 0.33 and x > 0.25 in the two cases, respectively. The integrated total number of electrons in the conduction band of LixMoS$_2$ are also shown for different values of x. This number matches the number of Li-ions in the cell in each case. PDOS plots showing gradual metallization of the upper MoS$_2$ layer upon increasing the concentration of intercalated Li-ion from 0.083 (c) to 0.333 (f).

While the above studies provide ample computational evidence that within a certain concentration of intercalated Li-ions, the system in its 2H phase can be properly photoexcited, we need to confirm that the 2H phase remains the most stable phase and does not undergo transitions to a metallic phase in this concentration regime. Previous studies have shown that while monolayer 1H-MoS$_2$ is more stable than 1T'-MoS$_2$,[30,42], the stability order switches upon Li-ion intercalation beyond a certain concentration as shown in Figure S4.[42] We wanted to investigate where the transition point lies in a type-II set-up. Moreover, previous studies have shown that the phase stability can change depending on the number of electrons present in the system. [42] Computing the energy of the two phases in an explicit heterojunction set-up, however, is not practically feasible since there is a significant lattice parameter mismatch between the 1T'-MoS$_2$ and MoO$_3$ structures.

In order to mimic the effect of the $MoO_3$ layers, we introduced a hypothetical layer of electronegative fluorine atoms on top of the lithiated $MoS_2$ surface (Figure 6a) to extract electron density from the CB of lithiated $MoS_2$ in either phase without affecting the band edges (Figure 6b). The amount of charge extracted by the fluorine sheet, obtained by integrating the differential PDOS, varies from 0.92e to 1.64e as the number of Li-ions is increased from 1 to 6 in a simulation cell with 12 units of $MoS_2$ (Table 1). The corresponding PDOS plots are shown in Figure S5. Using this set-up, we can compute the effect of the charge density extraction on the phase transition between the 2H- and the 1T'- phases upon lithiation. As shown in Figure 7, the transition point is shifted slightly towards higher lithium composition. This result implies that at a comparatively low concentration of intercalated Li-ions, the system will remain in the 2H phase that maintains a non-negligible band gap.

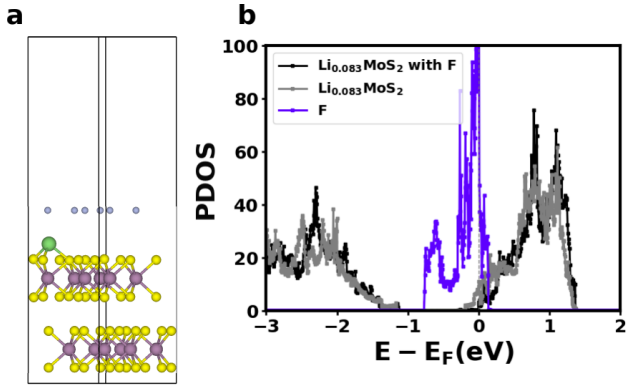

Figure 6. (a) Structure of $Li_{0.083}MoS_2$ + F6 sheet ; (b) Corresponding PDOS

Table 1. Number of electrons extracted by the F6 sheet for different Li-ion concentrations

| System | Charge Extracted by F-sheet |
| --- | --- |
| $Li_1Mo_{12}S_{24}/F_6$ | 0.920 e |
| $Li_2Mo_{12}S_{24}/F_6$ | 0.635 e |
| $Li_3Mo_{12}S_{24}/F_6$ | 1.601 e |
| $Li_4Mo_{12}S_{24}/F_6$ | 1.400 e |
| $Li_5Mo_{12}S_{24}/F_6$ | 1.321 e |
| $Li_6Mo_{12}S_{24}/F_6$ | 1.641 e |

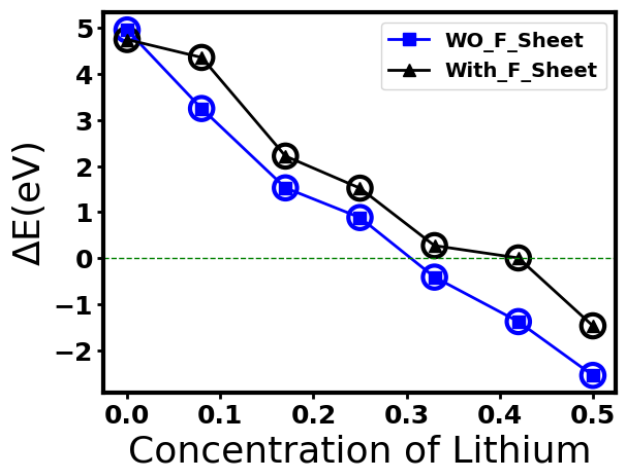

Figure 7. Phase transition with and without the hypothetical F6 sheet.

In order to experimentally investigate the effects of lithiation in the metallization and phase transition of monolayer (1H) MoS$_2$, we have systematically studied lithiation/delithiation using *in situ* Raman spectroscopy [details in the supporting information]. Pristine MoS$_2$ exhibits two prominent peaks at ~383 cm$^{-1}$ and ~402 cm$^{-1}$, which correspond to the E$^1_{2g}$ and A$_{1g}$ vibration modes of monolayer 1H-MoS$_2$, respectively. [41] On the other hand, the metallic 1T' phase is characterized by J1 and J2 modes centered at ~154 cm$^{-1}$ and ~225 cm$^{-1}$. [49,50] The scanning electron microscopy (SEM) image of monolayer MoS2 crystals transferred over an ITO coated quartz plate (ITO quartz) is shown in Figure 10a. A battery half-cell is constructed as explained before and as schematised in Figure 10b. Signature Raman peaks of the 1H and 1T' phases are shown in Figure 10c. The evolution of the vibrational modes of the 1H phase at different stages of charging is shown in Figure 10d.

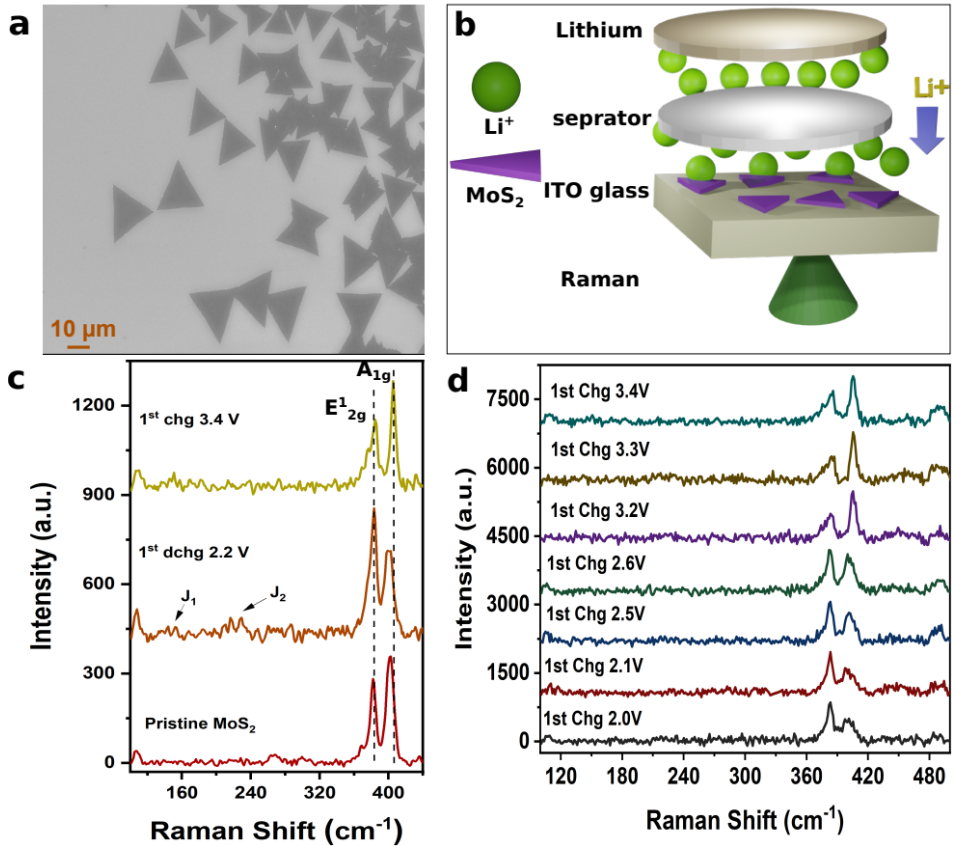

**Figure** 8. (a) SEM image of MoS$_2$ and (b) *the* in situ Raman set-up coupled with a Li ion battery half-cell. Raman in the schematic represents the transparent light entering-window of the cell helping for in situ studies. (c) The Raman spectra of MoS2 electrodes during lithiation/delithiation at different voltage *vs* Li/Li$^+$ of the pristine MoS$_2$ monolayer with an open circuit voltage of 2.8V, 1st discharge (dchg, 0.01 mAcm$^{-2}$) at 2.2V and after over charging (chg) till 3.4V. (d) The spectra taken during the electrochemical charging at different charging potentials.

It can be seen from the figure 10c that the pristine MoS2 has the respective Raman active phonons having a separation of ~ 19 cm-1 (E2g at 383 cm-1 and A1g at 402 cm-1), indicating the monolayer formation. Upon discharging the cell, Li ions will get attached to the MoS$_2$ and chemical binding via charge transfer will occur. The discharging (dchg) up to 2.2 V shows the bands corresponding to 1T' MoS$_2$, as discussed before. This indicates the incomplete phase transition of 1H MoS2. The electrochemical charging of the cell to 3.4 V is conducted further where the J1 and J2 phonon modes seem to have disappeared. Moreover, upon charging the cell the intensity ratio of E2g and A1g is also restored back to that of pristine indicating the deintercalation of the Li, in corroboration with the previous reports.[50]

Furthermore, we performed UV-Visible absorption studies for pristine MoS$_2$ and Li$_x$MoS$_2$ (discharged till 1V) to study the metal phase formation upon lithiation. As shown in Figure 11, pristine MoS$_2$ shows absorption peaks at 660 nm, 610 nm, and 434 nm corresponding to A, B, and C exciton of 1H MoS$_2$,[51] whereas Li$_x$MoS$_2$ does not show any absorption peak but higher absorption in the entire spectral range, indicating metallization upon lithiation.

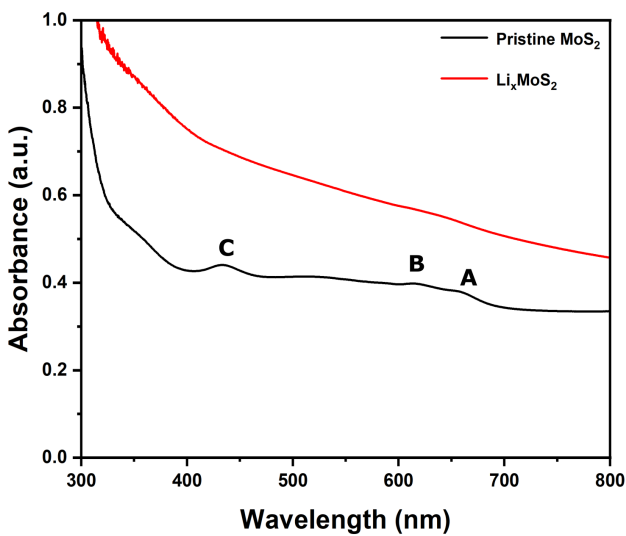

**Figure** 9: UV-Vis spectra of pristine and lithiated MoS$_2$.

   To understand the metallization of the 1H-phase upon lithiation, we performed *in situ* photoluminescence (PL) studies (Figure 12) at different discharge voltages. The drastic decrease in the intensity upon lowering the discharge voltage indicates decrease in efficiency of charge transfer from MoS$_2$ (lithiated) to MoO$_3$ and the shift in frequency to longer wavelengths imply decreasing band gap. Subsequent disappearance of photoluminescence below 2.95 V (during discharge) suggests disappearance of the type II character and phase transition to the 1T'-phase, which is also corroborated by the appearance of J1 and J2 peaks at that voltage in the Raman spectra. Detailed PL analyses at different voltages were performed and given in the supporting information indicating that reversible bandgap opening and closing can happen upon lithium insertion and de-insertion, following the 1T/1T' to 1H phase transition.

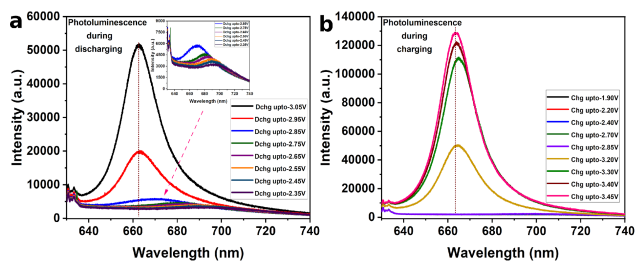

**Figure 10**: The *in situ* photoluminescence (PL) of MoS$_2$ monolayer based electrodes at different discharging voltages vs Li/Li$^+$ : (a) from charged state at 3.45 V to discharging till 2.35 V (inset PL curve shows discharge state from 2.85V to 2.35V). (b) during charging from discharged state 1.90 V to 3.45V.

**Conclusions**

In this paper, employing both computational and experimental techniques, we demonstrate that the type II heterojunction exposes an underlying band gap of the otherwise metallic Li$_x$MoS$_2$ but the band gap reduces upon increasing x and eventually the type II band alignment is lost (~ x > 0.25). The loss of the type II character would suppress the photo-cathode behavior since the photoexcited electron-hole pair cannot be separated efficiently anymore. Computationally, the transition from the 2H/1H phase to the 1T' phase is found to occur after the loss of the type II character. We believe that the dual role of the type II set-up (exposure of the underlying band gap and efficient formation of charge separated excitons upon photoexcitation) is a general phenomenon. On the other hand, the current study suggests that one should also consider the band structure modulation upon Li-ion intercalation as an important gauge for an ideal photocathode material. The band offset changes in MoS$_2$ upon lithium intercalation is rather drastic and hence photocathodes that are less prone to transitioning to a metallic phase are more desirable.

Recently it has been observed that the band offset changes in TiS$_2$ is negligible and hence the type II band setting can be preserved even with higher lithium concentration.[26] Hence, the photocathodes of higher capacity can be developed by the appropriate selection of materials that can open up new possibilities of next generation photo batteries. A systematic computational screening of such probable candidates is currently underway.

## ASSOCIATED CONTENT

**Supporting Information**

Experimental methods, additional plots, further details regarding band alignment

## AUTHOR INFORMATION

**Corresponding Authors**

**Soumya Ghosh** – *Tata Institute of Fundamental Research – Hyderabad, 36/P, Gopanpally Village, Serilingampally Mandal, Ranga Reddy District, Hyderabad 500046, India*; Orchid ID: orcid.org/0000-0002-9429-5238; Email ID: soumya.ghosh@tifrh.res.in

**Tharangattu N. Narayanan** - *Tata Institute of Fundamental Research – Hyderabad, 36/P, Gopanpally Village, Serilingampally Mandal, Ranga Reddy District, Hyderabad 500046, India*; Orchid ID: orcid.org/0000-0002-5201-7539; Email ID: tnn@tifrh.res.in

**Authors**

**Raheel Hammad** - *Tata Institute of Fundamental Research – Hyderabad, 36/P, Gopanpally Village, Serilingampally Mandal, Ranga Reddy District, Hyderabad 500046, India*

**Amar Kumar** - *Tata Institute of Fundamental Research – Hyderabad, 36/P, Gopanpally Village, Serilingampally Mandal, Ranga Reddy District, Hyderabad 500046, India*

**Notes**

There is no competing financial interesting

## ACKNOWLEDGEMENTS

The authors from TIFRH acknowledge the financial support from the Department of Atomic Energy, Government of India, under Project Identification No. RTI 4007. TNN and SG would like to acknowledge the funding support from Infosys-TIFR "Leading Edge" Research Grant. SG would also like to acknowledge SERB startup grant SRG/2021/001564

# Supplementary Information

**Photochargeable Li-ion Battery: Type II Heterojunction Exposes Underlying Band Gap in 'Metallic' Lithiated MoS$_2$**

*Raheel Hammad, Amar Kumar, Tharangattu N. Narayanan\*, and Soumya Ghosh\**

**This file includes:**

Section 1: Electronic structure calculations of Li$_x$MoS$_2$ and MoS$_2$/MoO$_3$ heterostructure and Phase transition of H-MoS$_2$ to T-MoS$_2$ in presence of Lithium and Fluorine Sheet

- Figure S1: Band structure calculations of 1H-MoS$_2$ for different Li concentrations

- Figure S2: Band edges of MoO$_3$(uL scheme) for different layers

- Figure S3: Bulk MoO$_3$ (uL scheme) and MoS$_2$(2L)offset

- Figure S4: Phase transition of 1H-MoS$_2$ to 1T'-MoS$_2$ upon Li intercalation

- Figure S5: PDOS calculations of Li$_x$MoS$_2$ for different X in presence of F sheet

Section 2: Experimental methods, characterization and Electrochemical measurement

- Figure S6: Schematic diagram of the MoS$_2$ growth by the CVD method

- Figure S7: In-situ Raman spectra of MoS$_2$ monolayer based electrodes during Lithiation/delithiation (with Lithium metal as counter electrode) at different voltage vs. Li/Li$^+$
.

## Section 1 Computational Methods:

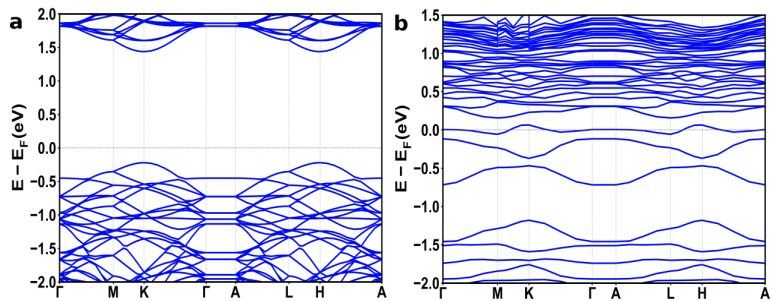

Figure S1: (a) Band structure of monolayer of 1H-MoS$_2$ computed with a periodic hexagonal cell with the lateral dimensions of 12.648 Å (16 Mo and 32 S atoms) employing PBE+U functional (b) Computed band structure of monolayer of 1H-Li$_{0.1875}$MoS$_2$

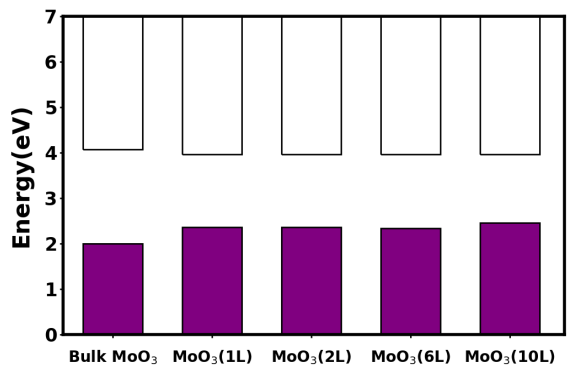

Figure S2: Change in the energy of the band edges on going for Different layers of MoO$_3$(uL scheme), computed using PBE+U+D3

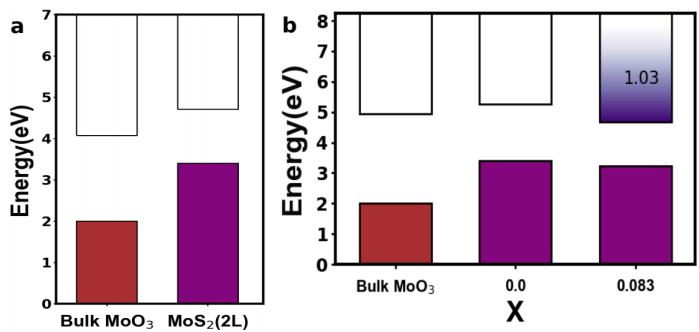

Figure S3. (a) Band gap is computed with HSE06 while the valence band offset computed with PBE+U within the uL scheme for MoO$_3$. (b) Change in band edges of lithiated MoS$_2$ as a function of Li-ion concentration. The type II alignment is lost for Li-ion concentration (x) less than 0.083

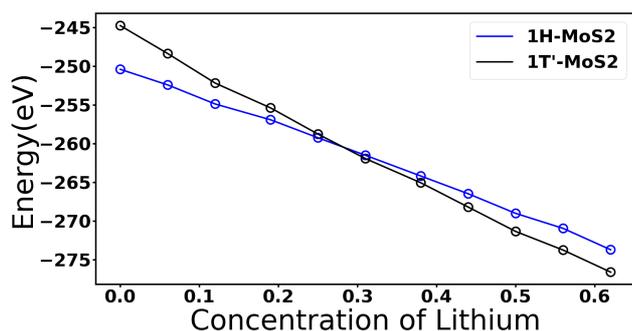

Figure S4. Phase transition in monolayer MoS$_2$ as a function of Li ion concentration for a periodic cell of 16 MoS$_2$ units. The simulation cell parameters for the hexagonal 1H and 1T' phases are 12.648 and 13.022 Å respectively. The energies are computed with PBE+U functional.

Integration of PDOS upto fermi level does not add up to total electrons in the system. To accurately compute the electron occupation number integration of total DOS has to be used. Incidentally, the PDOS of F and Li$_1$Mo$_{12}$S$_{24}$ (upto the Fermi level) do not overlap in Li$_1$Mo$_{12}$S$_{24}$/F$_6$ heterojunction. Therefore, the PDOS can be used to determine the integration limits and total DOS can be used to compute the electron occupations in flourine and Li$_1$Mo$_{12}$S$_{24}$. To compute electron occupations in Li$_x$Mo$_{12}$S$_{24}$/F$_6$ for x > 1, we compute the integral of PDOS for F upto fermi level for these systems. This integral is then compared with PDOS integral for F in Li$_1$Mo$_{12}$S$_{24}$/F$_6$, the difference is used to determine the charge extracted by flourine.

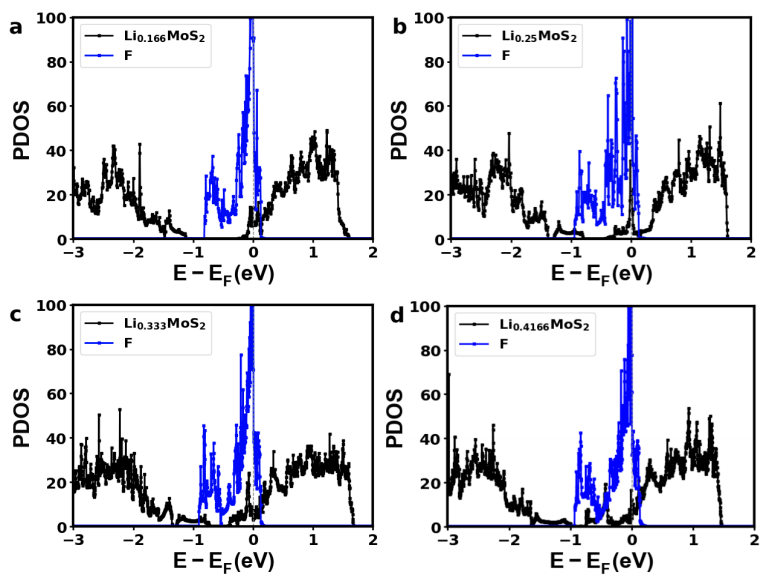

Figure S5. PDOS for different values of x in Li$_x$Mo$_{12}$S$_{24}$/F$_6$ system

**Section 2 Experimental Methods:**

The growth procedure for MoS$_2$ monolayer synthesis involved the use of chemical vapor deposition (CVD) method. A two-zone furnace was employed for the growth, following a method previously discussed. Only MoO$_3$ and sulfur were used as precursors, and a small amount of NaCl was added to aid metal decomposition at lower temperatures. The process involved keeping sulfur at 210°C in Zone I and placing MoO$_3$ at 710°C in Zone II, as indicated in the schematic diagram. The experiment was conducted in the presence of 190 sccm N$_2$, which acted as the carrier gas. Upon completion of the growth, the furnace was rapidly cooled to room temperature to prevent further multilayer growth.

**Characterization:**

The Renishaw Invia Raman spectroscopy was used to obtain in-situ and ex-situ Raman spectra and photoluminescence (PL) spectra. The analysis was conducted using a 532 nm exciton laser with a 20x objective. The laser power was optimized to prevent overheating and improve the noise to signal ratio of the data.

**Electrochemical measurements:**

Electrochemical measurements in this study were carried out using a single-channel Bio-logic potentiostat (SP-200). The electrochemical discharge measurements during the lithiation of monolayer MoS$_2$ were performed using a two-electrode battery setup consisting of an ITO-coated quartz electrode (25 mm × 25 mm) as the working electrode, lithium metal as the other electrode, and LiPF$_6$ in EC/EMC as the electrolyte. Prior to transferring the MoS$_2$ onto the ITO-coated quartz electrode, the electrode was thoroughly cleaned by washing it with soap and then with DI water and iso-propanol several times. The MoS$_2$ was first spin-coated with PMMA (0.204 g in 5 mL toluene) and then immersed in 2 M KOH overnight. The PMMA-coated MoS$_2$ was then transferred to water to remove KOH and subsequently transferred to the ITO-coated quartz electrode. The electrode was further cleaned with acetone after drying

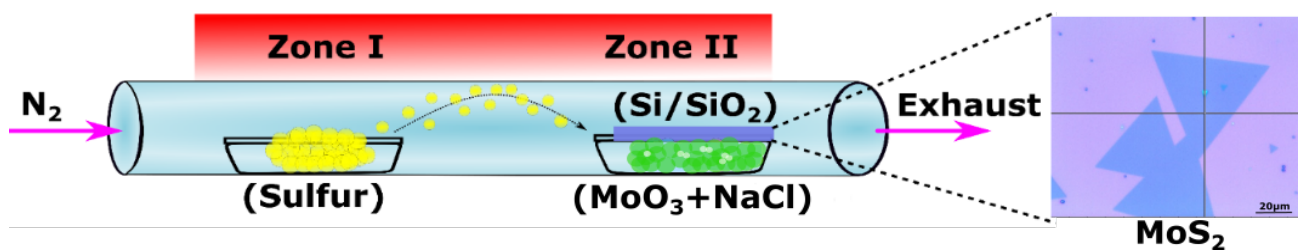

Figure S6. Schematic diagram of the MoS$_2$ growth by the CVD method. Sulfur is placed in Zone I whereas a mixture of MoO$_3$ and NaCl in Zone II. The MoS$_2$ on Si/SiO$_2$ is clearly seen in the optical image where the lateral size of the crystal is around 100-200 μm.

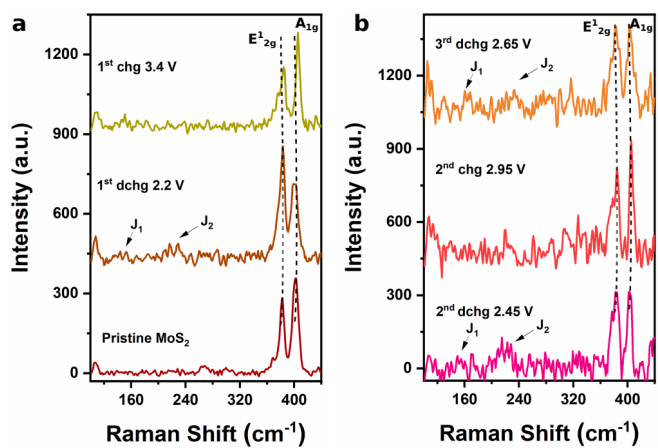

Figure S7. In-situ Raman spectra and Photoluminescence (PL) of $MoS_2$ monolayer based electrodes during Lithiation/delithiation (with Lithium metal as counter electrode) at different voltage vs. $Li/Li^+$. (a) Initial $MoS_2$ monolayer, 1st discharge at 2.2V and after charging till 3.4V (b) 2nd discharge at 2.45V, 2nd charge at 2.95V and 3rd discharge at 2.65V.